\documentclass{elsart}
\usepackage{epsfig}

\begin{document}

\begin{frontmatter}

\title{Comparing fragmentation of strange quark in 
{\mathversion{bold}$Z^0$} decays and {\mathversion{bold}$K^+p$} reactions}

\author{P.V. Chliapnikov}

\address{Institute for High Energy Physics, Protvino, RU-142284, 
        Russia\thanksref{mail}}
\thanks[mail]{E-mail address: chliapnikov@mx.ihep.su 
                                             (P.V. Chliapnikov)}

\begin{abstract}
{\scriptsize
The ratios of the production rates $\mathrm{K}^{*0}(892)/\mathrm{K}$, 
$\phi/\mathrm{K}$, $\rho^0/\pi$, $\omega/\pi$, $\Delta^{++}(1232)/\mathrm{p}$,
$\Sigma^{*+}(1385)/\Lambda$, $\Xi^-/\Lambda$ and their $x_p$-dependence 
obtained from results of the LEP and SLD experiments in $\mathrm{Z}^0$ 
hadronic decays are analyzed. The corresponding ratios for promptly produced 
mesons are estimated at $x_p \rightarrow 1$. A comparison of the LEP results 
with those from the Mirabelle and BEBC $\mathrm{K}^+p$ experiments at 32 and 
70 GeV/$c$ shows striking similarity in fragmentation of the $\bar s$ valence 
quark of the incident $\mathrm{K}^+$ and strange quarks produced in 
$\mathrm{Z}^0$ decays. The JETSET model describes the LEP, Mirabelle and
BEBC results. The model of Pei is consistent with the data for mesons, but 
presumably underestimates the fractions of primary octet baryons. The quark 
combinatorics model of Anisovich et al. is incompatible with the data.
}
\end{abstract}

\end{frontmatter}
%%%%%%%%%%%%%%%%%  For References  %%%%%%%%%%%%%%%%%%%%%%%%%%%%%%%%%%%%%%%%%
\def\zp{Z.\ Phys.\ {\bf C}}
\def\epj{Eur.\ Phys.\ J.\ {\bf C}}
\def\pl{Phys.\ Lett.\ {\bf B}}
\def\pr{Phys.\ Rev.\ {\bf D}}
\def\np{Nucl.\ Phys.\ {\bf B}}
\def\c {Collab.,\ }
\def\ac{ALEPH\ Collab.,\ }
\def\dca{DELPHI\ Collab.,\ P.\ Abreu\ et\ al.,\ }
\def\lc{L3\ Collab.,\ }
\def\oc{OPAL\ Collab.,\ }
%%%%%%%%%%%%%%%%%%%%%%%%%%%%%%%%%%%%%%%%%%%%%%%%%%%%%%%%%%%%%%%%%%%%%%%%%%%%%
\def\beq{\begin{equation}}
\def\eeq{\end{equation}}
\def\bear{\begin{eqnarray}}
\def\enar{\end{eqnarray}}
\def\nnb{\nonumber}
\def\nin{\noindent}
%%%%%%%%%%%%%%%%  General Symbols %%%%%%%%%%%%%%%%%%%%%%%%%%%%%%%%%%%%%%%%%%%%
\newcommand \ee {\ifmmode e^+e^-    \else $e^+e^-$\fi}
\newcommand \zo {\ifmmode \mathrm{Z^0} \else $\mathrm{Z^0}$\fi}
\def\zss{$\mathrm{Z^0} \rightarrow s\bar{s}$}
\newcommand \pv{\ifmmode P_V \else $P_V$\fi}
\newcommand \ps{\ifmmode P^s_V \else $P^s_V$\fi}
\def\la{\langle}
\def\ra{\rangle}
\def\ol{\overline}
\newcommand \sfac{\ifmmode 2J +1     \else $2J +1$\fi}
%===================  Mesons  =============================%
\newcommand\kpp{\ifmmode \mathrm{K}^+p    \else $\mathrm{K}^+p$\fi}
\newcommand \kl{\ifmmode \mathrm{K}       \else $\mathrm{K}$\fi}
\newcommand \ko{\ifmmode \mathrm{K}^0     \else $\mathrm{K}^0$\fi}
\newcommand \ako{\ifmmode \mathrm \bar{K}^0 \else $\bar{\mathrm{K}}^0$\fi}
\newcommand \kp{\ifmmode \mathrm{K}^+     \else $\mathrm{K}^+$\fi}
\newcommand \km{\ifmmode \mathrm{K}^-     \else $\mathrm{K}^-$\fi}
\newcommand \kv{\ifmmode \mathrm{K}^*     \else $\mathrm{K}^*$\fi}
\newcommand \kvo{\ifmmode \mathrm{K}^{*0} \else $\mathrm{K}^{*0}$\fi}
\newcommand \kvp{\ifmmode \mathrm{K}^{*+} \else $\mathrm{K}^{*+}$\fi}
\newcommand \kvn{\ifmmode \mathrm{K}^*(892)\else $\mathrm{K}^*(892)$\fi}
\newcommand \kvon{\ifmmode \mathrm{K}^{*0}(892)\else $\mathrm{K}^{*0}(892)$\fi}
\newcommand \pis{\ifmmode \pi             \else $\pi$\fi}
\newcommand \pip{\ifmmode \pi^+           \else $\pi^+$\fi}
\newcommand \piz{\ifmmode \pi^0           \else $\pi^0$\fi}
\newcommand \pim{\ifmmode \pi^-           \else $\pi^-$\fi}
\newcommand \roz{\ifmmode \rho^0          \else $\rho^0$\fi}
\newcommand \oms{\ifmmode \omega          \else $\omega$\fi}
\newcommand  \ph{\ifmmode \phi            \else $\phi$\fi}

%%%%%%%%%%%%%%%%%%       Baryons       %%%%%%%%%%%%%%%%
\newcommand \prot{\ifmmode \mathrm{p}           \else $\mathrm{p}$\fi}
\newcommand \del{\ifmmode \Delta^{++}           \else $\Delta^{++}$\fi}
\newcommand \deln{\ifmmode \Delta^{++}(1232)    \else $\Delta^{++}(1232)$\fi}
\newcommand  \lam{\ifmmode \Lambda              \else $\Lambda$\fi}
\newcommand \alam{\ifmmode \bar \Lambda         \else $\bar \Lambda$\fi}
\newcommand \sigo{\ifmmode \Sigma^0             \else $\Sigma^0$\fi}
\newcommand \asigo{\ifmmode \ol{\Sigma^0}       \else $\ol{\Sigma^0}$\fi}

\newcommand \sigr{\ifmmode \Sigma^*             \else $\Sigma^*$\fi}
\newcommand \sigrn{\ifmmode \Sigma^*(1385)      \else $\Sigma^*(1385)$\fi}
\newcommand \sigrp{\ifmmode \Sigma^{*+}         \else $\Sigma^{*+}$\fi}
\newcommand \sigrpn{\ifmmode \Sigma^{*+}(1385)  \else $\Sigma^{*+}(1385)$\fi}
\newcommand \asigr{\ifmmode \ol{\Sigma}^*       \else $\ol{\Sigma^*}$\fi}
\newcommand\asigrn{\ifmmode \ol{\Sigma^*}(1385) \else $\ol{\Sigma^*}(1385)$\fi}
\newcommand \asigrp{\ifmmode \ol{\Sigma^{*+}}   \else $\ol{\Sigma^{*+}}$\fi}
\newcommand \asigrm{\ifmmode \ol{\Sigma^{*-}}   \else $\ol{\Sigma^{*-}}$\fi}
\newcommand \xim{\ifmmode \Xi^-                 \else $\Xi^-$\fi}
\newcommand \axim{\ifmmode \ol{\Xi^-}           \else $\ol{\Xi^-}$\fi}

The vector and pseudoscalar mesons, or decuplet and octet baryons, 
differ in the relative orientation of the quark spins. Therefore the 
ratios of their production rates, $V/P$ and $D/O$, provide us with 
important information on spin dependence in fragmentation. 
Hadronization models predict $V/P$ and $D/O$ for promptly produced 
particles, not resulting from decays of other particles or resonances. 
Therefore apart from the models based on Monte-Carlo generators, 
such as the JETSET model \cite{jetset}, these predictions are difficult 
to test experimentally.

In the nonrelativistic quark model, the mass difference of the vector 
and pseudoscalar mesons, decuplet and octet baryons is explained by the 
hyperfine mass splitting. The production rates exhibit a strong mass 
dependence. Therefore the vector-to-pseudoscalar and decuplet-to-octet 
suppressions could also be explained, at least qualitatively, by the 
hyperfine mass splitting \cite{h_1,h_2} provided that the rates of 
promptly produced particles were reliably estimated.
 
The simplest spin model of fragmentation, such as the model suggested 
in \cite{as} and subsequently developed by Anisovich and his 
collaborators in many other papers, is a random combination of quark 
spin states giving $V/P = 3$ and $D/O = 2$. Recently Anisovich et al. 
\cite{ann} compared these predictions with the LEP data. They analyzed 
the $V/P$ ratios obtained from the ALEPH data \cite{Aleph} at large 
$x_p = p/p_{beam} \rightarrow 1$, where a contribution of the resonance 
decays is supposed to be strongly suppressed, and concluded that the 
experimental results are consistent with $V/P = 3$. This conclusion is 
in contradiction with the results obtained in \cite{h_1,h_2}. Besides, 
the LEP results on hadron production, in general, and ALEPH results 
\cite{Aleph}, in particular, agree well with the JETSET model 
\cite{jetset}. In JETSET, the probability to produce a vector meson is 
controlled by the parameter $V/(V+P)$. This parameter pertaining to 
mesons directly produced in hadronization is much smaller than 0.75, 
the value expected from the quark-combinatorics prediction. 

In order to resolve these problems and to get better insight into the
nature of the vector-to-pseudoscalar and decuplet-to-octet suppressions
we investigate in this paper the ratios of the production rates \footnote
{Apart from Fig.~1, the charge conjugates and antiparticles are not 
included into definition of the rates in this paper.} \kvn/\kl, \ph/\kl, 
\roz/\pis, \oms/\pis, \deln/\prot, \sigrn/\lam\ and \xim/\lam\  obtained 
from results of the LEP [6-17] and SLD \cite{SLD} experiments. We estimate 
the values of these ratios for promptly produced hadrons by studying 
their $x_p$-dependence and comparing the fragmentation of strange quark 
in \zo\ decays and \kpp\ reactions. Indeed, a relatively strong 
suppression of multi-strange final states in kaon induced reactions at 
moderate energies offers good possibilities to trace the flow of the 
incident strange valence flavour among the reaction debris. The \ko, 
\kvon, \ph, \alam, \asigrn\ and \axim\ in the Mirabelle [19-24] and 
BEBC [25-27] \kpp\ experiments at 32 and 70 GeV/$c$ are dominantly 
produced on the $\bar s$ valence quark of the incident \kp. This is 
clearly seen from their Feynman-$x_F$ spectra exhibiting a prominent 
leading particle effect and, consequently, from a relatively small 
production of these particles in the central region. This allows to 
obtain the reliable estimates of $V/P$ and $D/O$ ratios for promptly 
produced hadrons. Therefore an interesting and sensible test of the 
LEP results, as well as model predictions, on $V/P$ and $D/O$ ratios 
for promptly produced hadrons is feasible with kaon beams.

In studying the ratios \kvn/\kl, \ph/\kl, \roz/\pis, \oms/\pis, 
\deln/\prot, \sigrn/\lam\ and \xim/\lam, the differential cross-sections, 
$1/\sigma_h\cdot d\sigma/dx_p$, for the resonances and \xim\ were taken 
from the LEP experiments [6-17]. In some of them, the inclusive spectra 
have been presented as a function of $x_E = E/E_{beam}$. The difference 
between $x_p$ and $x_E$, important only at very small values of $x_p$ 
and $x_E$, was ignored. The differential cross-sections for the \kl, \pis, 
\prot\ or \lam\ in the same $x_p$-intervals were taken either from the 
same experiments or from combined $x_p$-spectra of corresponding particles 
measured by ALEPH \cite{Aleph}, DELPHI \cite{Dkpip}, OPAL \cite{OKpip} and 
SLD \cite{SLD} and compiled in \cite{biebel}. These spectra were fitted 
by a sum of two exponentials, and cross-sections in the corresponding 
$x_p$-intervals were calculated using the results of these fits. Such 
combined $x_p$-spectra for the $\kl^{\pm}$ and $\pi^{\pm}$ with the 
results of their fits by exponentials are illustrated, for example, 
in Fig.~1.
\begin{figure}[tbhp]
\noindent
\begin{minipage}{0.50\linewidth}
\includegraphics[width=\linewidth]{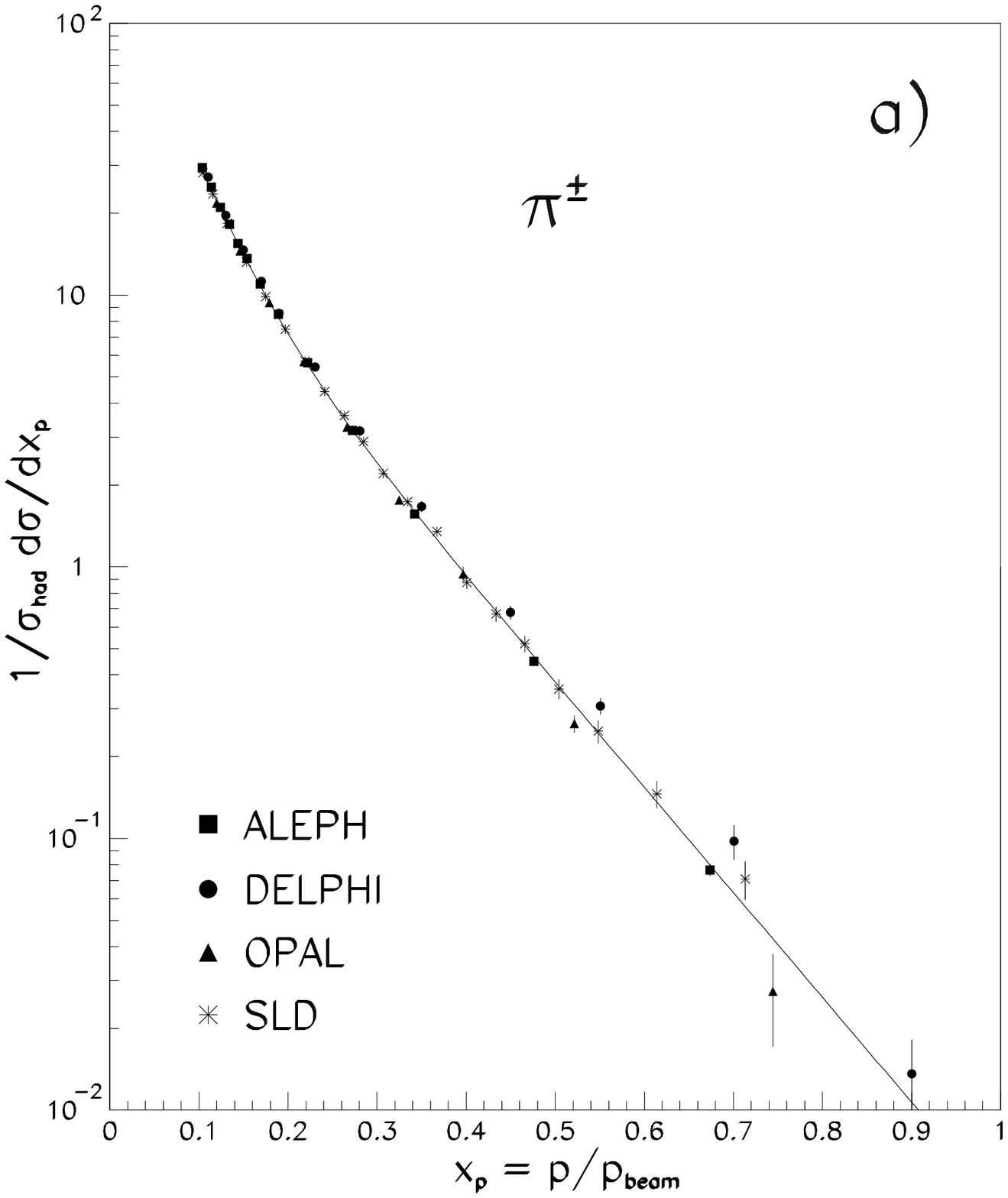}
%\centering{\epsfig{figure=pispectra_bl.eps,width=\linewidth}}
\end{minipage}\hfill
\begin{minipage}{.50\linewidth}
\includegraphics[width=\linewidth]{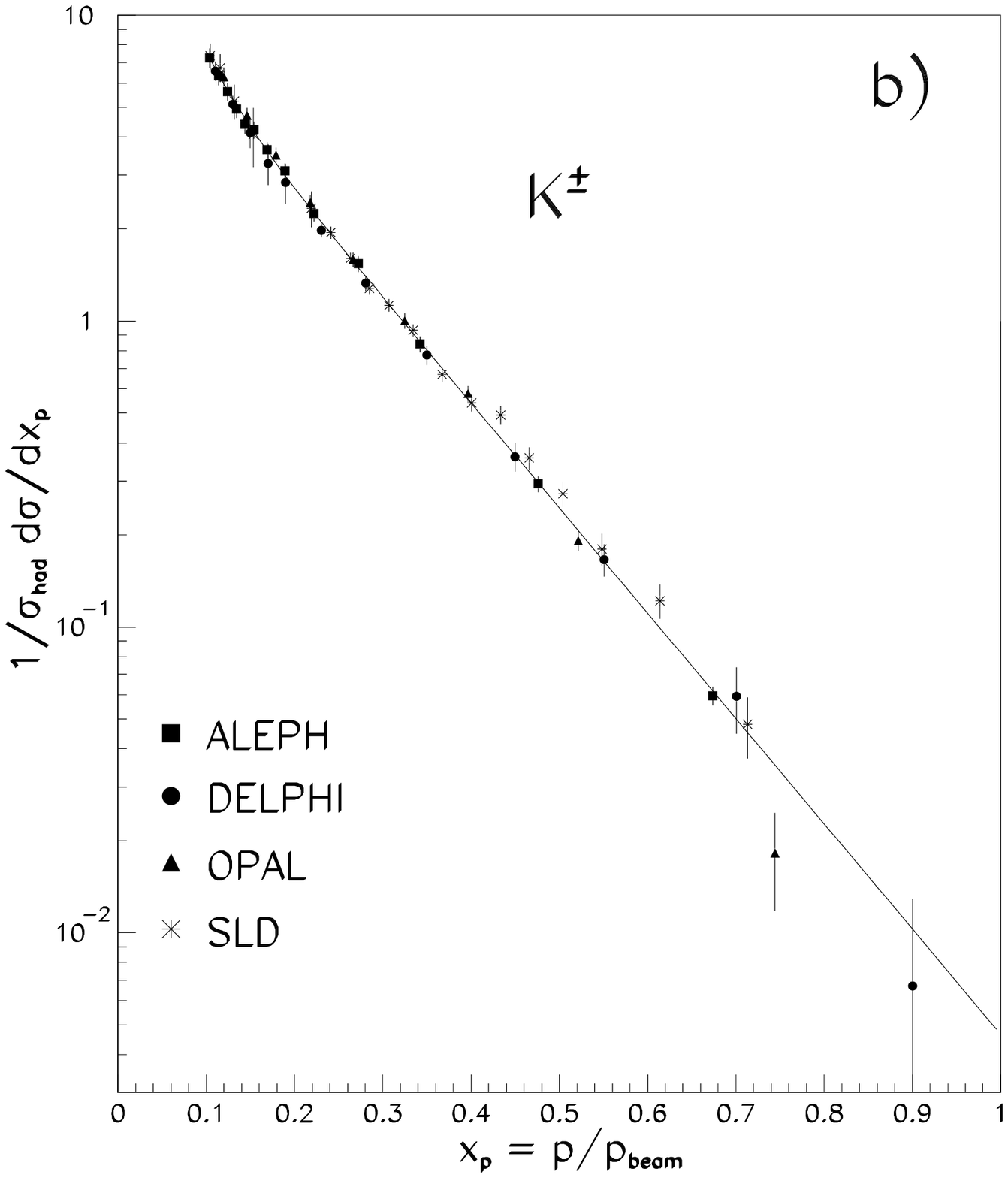}
%\centering{\epsfig{figure=kspectra_bl.eps;1,width=\linewidth}}
\end{minipage}\hfill
\caption{\scriptsize The $\pi^{\pm}$ (a) and $\kl^{\pm}$ (b) differential 
    cross-sections, $1/\sigma_h\cdot d\sigma/dx_p$, for $x_p > 0.1$ 
    measured by ALEPH, DELPHI, OPAL and SLD. The {\it curves} are the 
    results of the fit by a sum of two exponentials.}
\end{figure}

The \kvon/(3\kp) ratios as a function of $x_p$ are presented in Fig.~2a. 
The \kvon\ differential cross-sections were taken from the ALEPH 
\cite{Aleph}, DELPHI \cite{Dkvophi} and OPAL \cite{Okv} experiments. 
The results of three LEP experiments are well consistent. The \kvon/(3\kp) 
ratio increases with $x_p$, presumably approaching the value for the
promptly produced mesons at $x_p \rightarrow 1$. The ratios 
$\kvon_{prompt}/(3\ko_{prompt}) = 0.33 \pm 0.06$ and $0.28 \pm 0.04$ 
for the promptly produced mesons measured in the Mirabelle \cite{Mkvtokphi} 
and BEBC \cite{Bkvtok} \kpp\ bubble chamber experiments at 32 and 
respectively 70 GeV/$c$ are also shown in Fig.~2a\footnote{The BEBC value 
was corrected for the \ako\ production cross-section \cite{Bantiko} not 
accounted for in \cite{Bkvtok}.}. They are in excellent agreement with 
a trend of the LEP data for $x_p \rightarrow 1$. They are also well 
consistent with the predictions of the JETSET \cite{jetset}\footnote{Here 
and below the JETSET 7.4 predictions for the total rates given in 
\cite{Aleph} were used. The fractions of the directly produced particles 
in JETSET were taken from \cite{pei}.} and Pei \cite{pei} models for the 
promptly produced mesons at LEP also shown in Fig.~2a. On the other hand, 
and contrary to conclusion of ref. \cite{ann} based on the same analysis 
of the ALEPH data, the quark-combinatorics prediction, \kvn/(3\kl) = 1, 
for the promptly produced mesons is ruled out by the LEP, Mirabelle and 
BEBC data. The experimental values at the largest $x_p$ are a factor of 3 
lower than the prediction and deviate from it by at least 11 standard 
deviations. 
\begin{figure}[th]
\noindent
\begin{minipage}{0.48\linewidth}
\includegraphics[width=\linewidth]{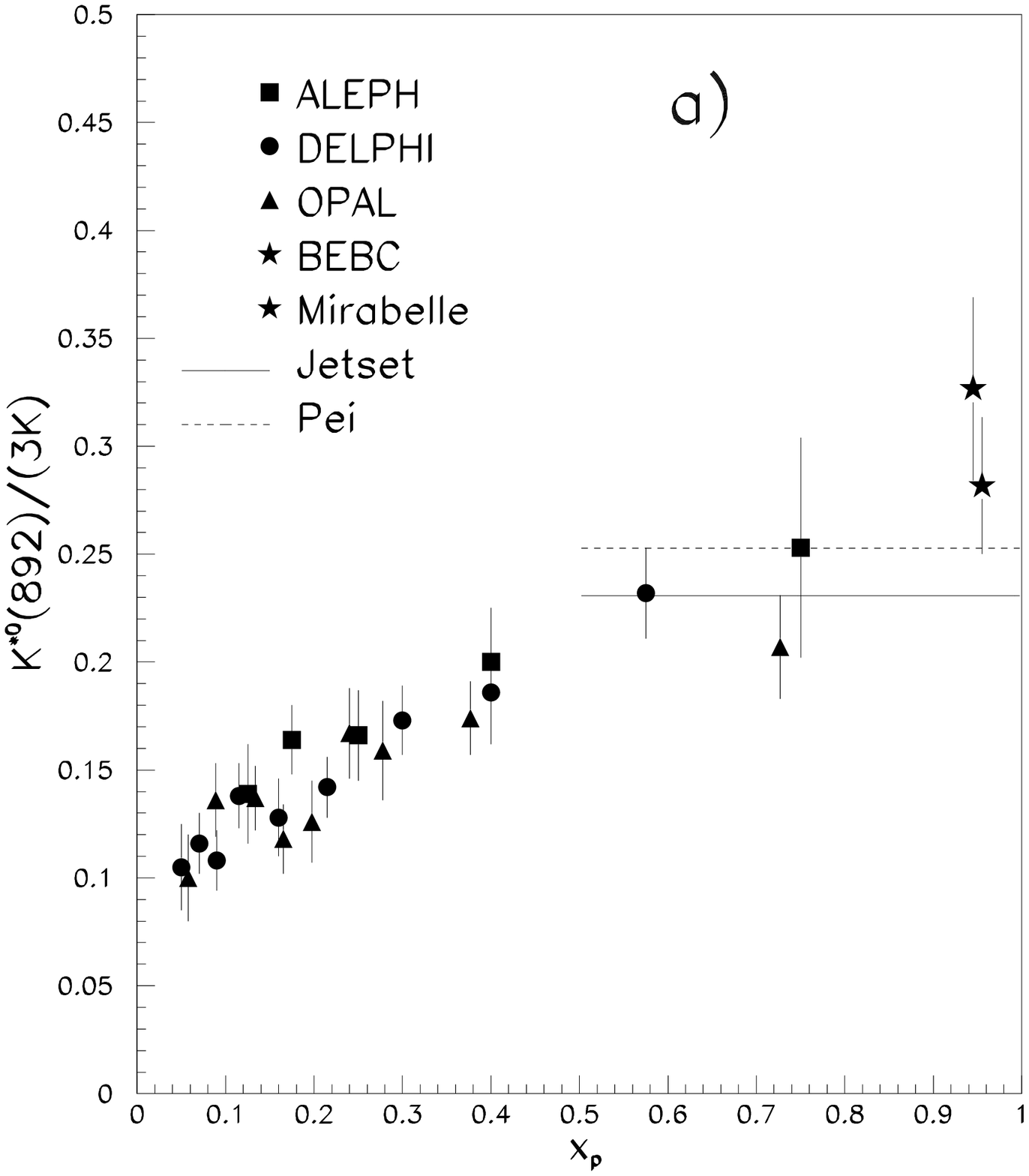}
%\centering{\epsfig{figure=kvtok_bl.eps,width=\linewidth}}
\end{minipage}\hfill
\begin{minipage}{.48\linewidth}
\includegraphics[width=\linewidth]{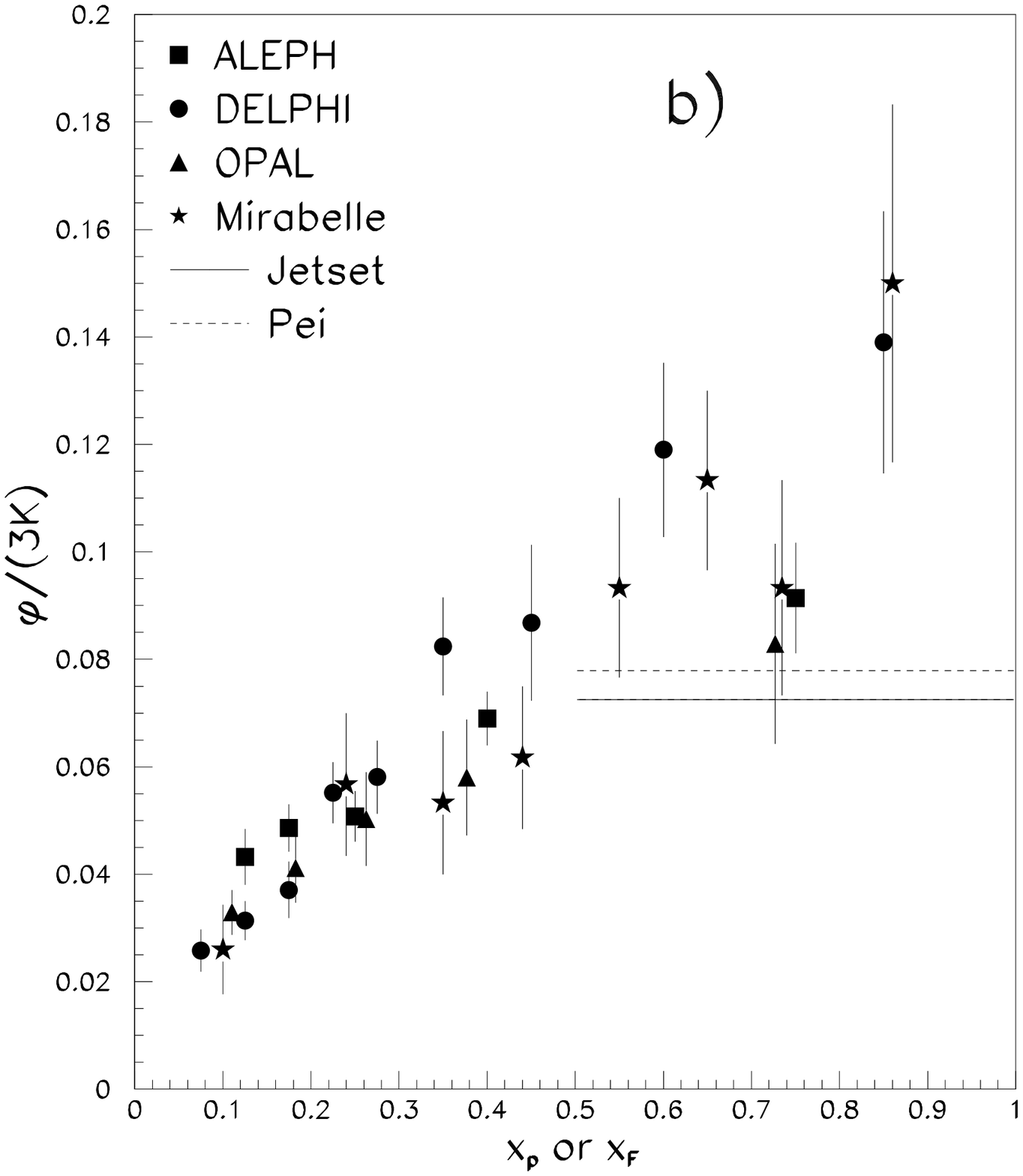}
%\centering{\epsfig{figure=phitok_bl.eps;1,width=\linewidth}}
\end{minipage}\hfill
\caption{\scriptsize The \kvon/(3\kp) (a) and \ph/(3\kp) (b) ratios
    obtained from 
    the LEP experiments as a function of $x_p$ ; the \kvon/(3\ko) ratios for 
    the promptly produced particles from the Mirabelle and BEBC \kpp\ 
    experiments at 32 and 70 GeV/$c$ (a); the $\ph/(3\ko_s)$ ratio as 
    a function of Feynman-$x_F$ from Mirabelle (b). The JETSET and Pei 
    model predictions for the promptly produced particles at LEP are also 
    shown. The quark-combinatorics model predicts $\kvn/(3\kl) = 1$ and 
    $\ph/(3\kl) = \lambda$. Here and in the subsequent figures, some data 
    points were slightly shifted to avoid overlap.}
\end{figure}

The \ph/(3\kp) ratios, with the \ph\ differential cross-sections 
measured by ALEPH \cite{Aleph}, DELPHI \cite{Dkvophi} and OPAL 
\cite{Ophi}, are presented as a function of $x_p$ in Fig.~2b. 
The data from three LEP experiments are consistent within errors 
and exhibit a clear rise with increasing $x_p$, very similar to 
the behaviour seen in Fig.~2a. 
The $\ph/\ko_s$ ratio as a function of Feynman variable $x_F = 
p_L/(\sqrt{s}/2)$, where $p_L$ is the longitudinal momentum in 
the centre-of-mass system, from Mirabelle \cite{Mrhophi} is also 
presented in Fig.~2b for $x_F > 0$. It is strikingly similar to 
the behaviour of the LEP data. The JETSET and Pei model predictions are 
somewhat lower than the experimental data at large $x_p$, but consistent 
with them within one-two errors. The quark-combinatorics model 
\cite{as,ann} predicts $\ph/(3\kl) = \lambda$, where $\lambda$ is 
the strangeness suppression parameter. For usually accepted value 
$\lambda \approx 0.3$, this prediction is a factor of more than 
two higher than the experimental values at large $x_p$. It deviates 
from the DELPHI value of $0.14 \pm 0.02$ at $x_p = 0.85$ 
by 7 standard deviations or even more for $\lambda \approx 0.5$ 
suggested in \cite{ann}.  
\begin{figure}[tbhp]
   \begin{center} 
  \epsfig{file=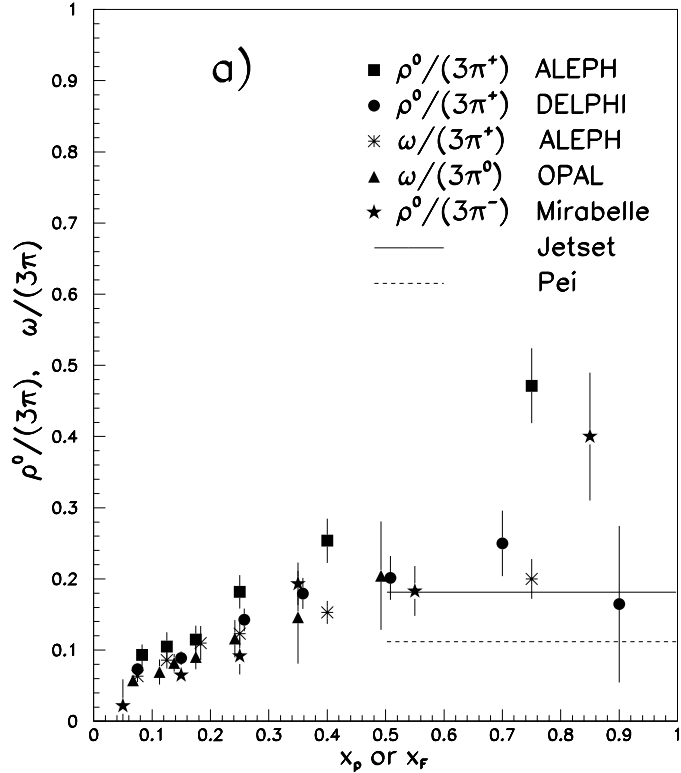,width=10.0cm}
   \end{center}
  \caption{\scriptsize The \roz/(3\pip), \oms/(3\pip) and \oms/(3\piz) ratios 
    obtained from the LEP experiments as a function of $x_p$, together with 
    the $\roz/(3\pim)$ ratio from the Mirabelle \kpp\ experiment at 32 GeV/$c$
    as a function of Feynman-$x_F$. The \roz/(3\pip) ratios for the promptly 
    mesons at LEP predicted by the JETSET and Pei models are 
    also shown. The quark-combinatorics model predicts  $\roz/(3\pi) = 1$.}
\end{figure}
 
The \roz/(3\pip) ratios, with the \roz\ differential cross-sections 
measured by ALEPH \cite{Aleph} and DELPHI \cite{Droz}, are presented 
as a function of $x_p$ in Fig.~3. The results of these experiments 
are consistent within errors for $x_p < 0.5$.  For $x_p> 0.5$, the 
ALEPH data point is significantly higher than the DELPHI data points.
The \oms/(3\pis) ratios measured by ALEPH \cite{Aleph} and OPAL 
\cite{Ooms} are also presented in Fig.~3. They are consistent with 
the \roz/(3\pip) ratios for $x_p < 0.5$. For $x_p > 0.5$, the ratio 
\oms/(3\piz) measured by ALEPH is close to the \roz/(3\pip) ratio 
measured by DELPHI. The \roz/(3\pim) ratio as a function of 
Feynman-$x_F$ taken from the Mirabelle experiment \cite{Mrhophi} 
(and convoluted around $x_F = 0$) is also presented in Fig.~3. 
It is well consistent with the LEP results for $x_p < 0.7$ and lies 
between the ALEPH and DELPHI data points at $x_p$ = 0.7-1.0. Thus, 
in spite of some inconsistency of the LEP results at the largest $x_p$, 
the general tendency of the LEP and Mirabelle data in Fig.~3 is 
quite similar to the one observed in Fig.~2. However, in this case 
the JETSET and especially Pei model predictions for the ratios of 
promptly produced mesons at LEP appear to be underestimated. The 
quark-combinatorics prediction, $\roz/(3\pis) = 1$, is not consistent 
with the data, contrary to conclusion of ref. \cite{ann} based on 
the same analysis of the ALEPH data. It is again a factor of 2 to 5 
higher than the experimental values at large $x_p$. The values 
$\roz/(3\pip) = 0.250 \pm 0.046$ at $x_p$ = 0.6-0.8 \cite{Droz} and 
$0.47 \pm 0.05$ at $x_p$ = 0.5-1.0 \cite{Aleph} deviate from this 
prediction by 16 and respectively 11 standard deviations.

In principle, the experimental values of the vector-to-pseudoscalar
ratios in the fragmentation region at $x_p \rightarrow 1$ could be 
reconciled with the quark combinatorics predictions in case of strong 
spin alignment of vector mesons, since in this case the expected 
value would be $V/P = 1$. The spin density matrix elements for the 
\roz, $\rho^{\pm}$, $\omega$, \ph\ and \kvon\ have been measured by 
DELPHI \cite{Dsa} and OPAL [31-33] (a summary of the results can be 
found in \cite{Osarozom}). The helicity density matrix elements 
$\rho_{00}$ of the \roz, $\rho^{\pm}$ and $\omega$ are found compatible, 
over the entire $x_p$ range, with 1/3 corresponding to a statistical 
mix of helicity -1, 0 and +1 states, with no evidence for spin alignment. 
For the \kvon\ and \ph\ produced in the small $x_p$ region $x_p \leq 0.3$, 
the values of $\rho_{00}(\kvo) = 0.33 \pm 0.05$ and $\rho_{00}(\ph) = 
0.30 \pm 0.04$ \cite{Dsa} are also well consistent with no spin alignment. 
For the \ph\ produced in the high $x_p \geq 0.7$ region, some indication 
on unequal population in the three helicity states is observed, with 
$\rho_{00} = 0.55 \pm 0.10$ \cite{Dsa} and $0.54 \pm 0.08$ \cite{Osaph}. 
For the \kvo, DELPHI \cite{Dsa} measured $\rho_{00} = 0.46 \pm 0.08$ at 
$x_p \geq 0.4$ and OPAL \cite{Osakv} obtained $\rho_{00} = 0.66 \pm 0.11$ 
at $x_p \geq 0.7$. It is interesting that the LEP results on the \kvo\ 
spin alignment are again very similar to the results of the Mirabelle 
\kpp\ experiment \cite{Msakv}. Thus in spite of some preference for 
occupation of the helicity zero state observed for the \kvo\ and \ph\
at large $x_p$, this can not explain the failure of the quark combinatorics 
model.
\begin{figure}[tbhp]
   \begin{center} 
  \epsfig{file=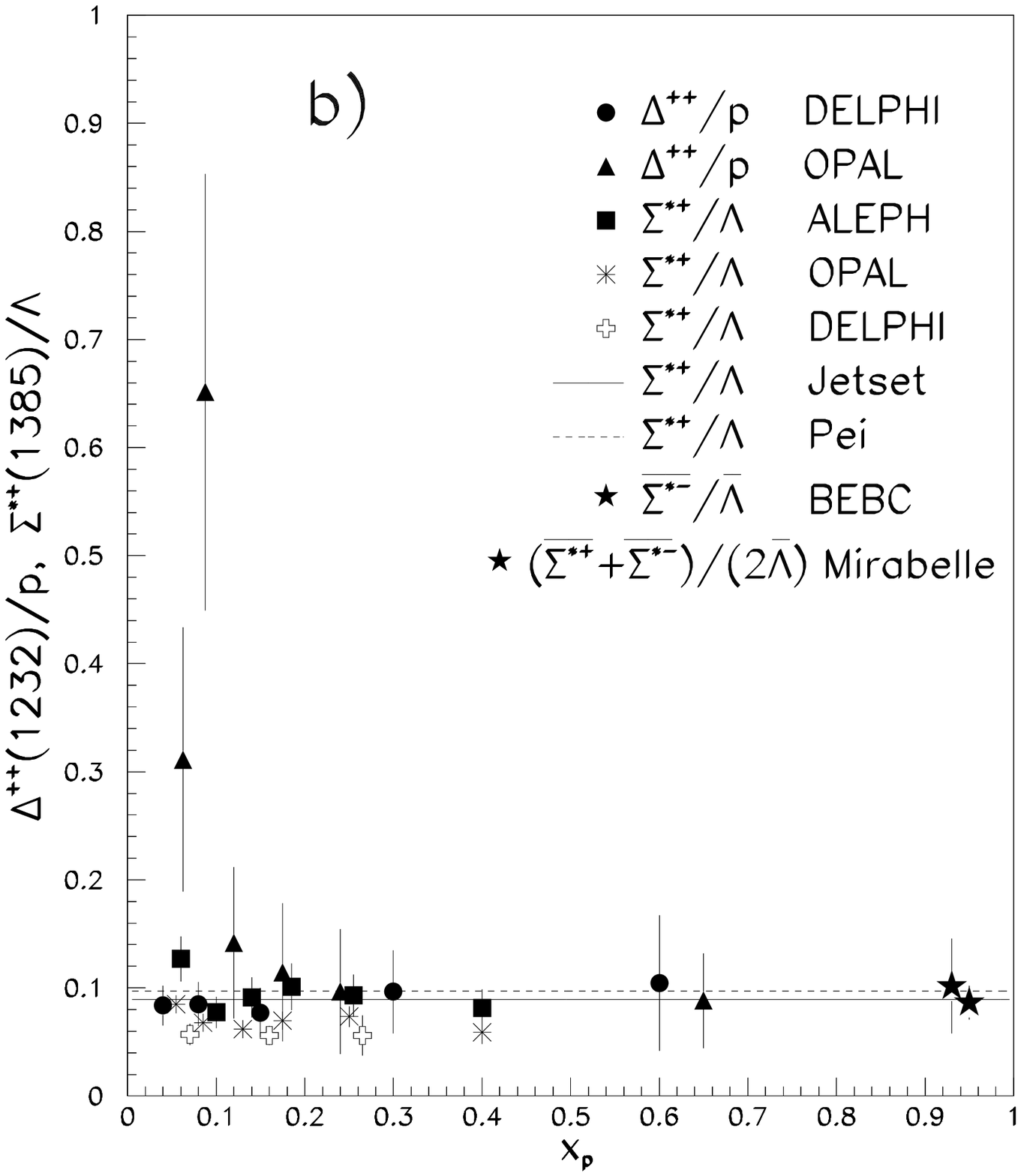,width=10.0cm}
   \end{center}
  \caption{\scriptsize The \deln/\prot\ and \sigrpn/\lam\ ratios as a function 
    of $x_p$ obtained from the LEP experiments and the \sigrpn/\lam\ ratios 
    predicted by the JETSET and Pei models for the total rates at LEP. 
    The ratios of the total rates $(\asigrp + \asigrm)/(2\alam)$ from  
    Mirabelle and \asigrm/\alam\  from BEBC are also shown.}
\end{figure}

The kinematics of the decuplet baryon decays does not allow to obtain 
information about ratios of promptly produced decuplet and octet 
baryons at large $x_p$ as have been possible for mesons. Still a 
study of $x_p$-dependence of the decuplet-to-octet ratios allows to 
make several important observations. The \deln/\prot\ ratio as a 
function of $x_p$ is presented in Fig.~4. The \deln\ differential 
cross-section was measured by DELPHI \cite{Ddel} and OPAL \cite{Odel}. 
The $x_p$-dependence of the \deln/\prot\ ratio obtained from the DELPHI 
measurements is essentially flat, as expected. The OPAL results are 
well consistent with those of the DELPHI for $x_p > 0.1$, but show 
quite unexpected behaviour at the smallest $x_p$. The discrepancy 
in the \deln\ total rate in two experiments is obviously due to 
those suspicious OPAL data points at the smallest $x_p$, where the 
extraction of the \deln\ signal is the most difficult. This is 
supported by comparison of $x_p$-dependences for the \deln/\prot\ and 
\sigrpn/\lam. The latter is also presented in Fig.~4. The data for 
the \sigrpn\ were taken from ALEPH \cite{Aleph}, DELPHI \cite{DXiSigr} 
and OPAL \cite{OlamSigrXi}. The \sigrpn/\lam\ ratio as a function of 
$x_p$ is essentially flat. This behaviour is very similar to the one 
observed for the \deln/\prot\ ratio by DELPHI in all $x_p$ range and 
by OPAL for $x_p > 0.1$. The ratios of the total rates 
\beq
(\asigrp + \asigrm)/(2\alam) = 0.086 \pm 0.016,~~~  \asigrm/\alam \ = 
                                                    0.102 \pm 0.044
\eeq
obtained respectively from Mirabelle \cite{Mbar,Mstr} and BEBC 
\cite{Bbar} (and also shown in Fig.~4) agree very well with the LEP 
results for the \sigrp/\lam\ and DELPHI results for the \del/\prot. 
The JETSET and Pei model predictions for the ratio \sigrp/\lam\ of 
total rates at LEP are also indicated in Fig.~4. They are well 
consistent with the LEP, Mirabelle and BEBC data.

Finally we compared the ratios \xim/\lam\ obtained from the LEP 
\cite{Aleph,DXiSigr,OlamSigrXi}, Mirabelle \cite{Mstr,Maxim} and 
BEBC \cite{Bbar} experiments. With the \xim\ and \lam\ total rates 
averaged over results of all LEP experiments \cite{h_1}, one gets 
$(\xim/\lam)_{tot} = 0.072 \pm 0.004$. From the \axim\ \cite{Maxim} 
and \alam\ \cite{Mstr} total rates in Mirabelle, one obtains 
$(\axim/\alam)_{tot} = 0.086 \pm 0.021$, in good agreement with 
the LEP value. The corresponding value, $0.14 \pm 0.04$, from 
BEBC \cite{Bbar} is consistent within errors with the LEP and 
Mirabelle results. The JETSET and Pei model predictions for the 
ratio of the total rates at LEP, 0.090 and 0.072, respectively, 
are in good agreement with the LEP, Mirabelle and BEBC results.

In view of good agreement between the LEP and bubble chamber 
experiments on the ratios \sigrp/\lam\ and \xim/\lam\ of the total 
rates, it is reasonable to assume that the corresponding ratios 
for the promptly produced baryons in these experiments might also 
be the same. The \asigrp, \asigrm\ and \axim\ in the Mirabelle 
\kpp\ experiment at 32 GeV/$c$ can presumably be safely considered 
as promptly produced. Indeed, even at LEP, the fractions of promptly 
produced \sigrp\ and \xim\ equal 0.92 and 0.75 in the JETSET, and 
0.91 and 0.41 in the Pei models. The cross-section of the promptly 
produced \alam\ in Mirabelle can be estimated from the \alam\ total 
rate \cite{Mstr} after subtraction of \alam\ from the \asigrp, 
\asigrm\ and $\overline{\Sigma^{*0}}$ decays\footnote{Assuming that 
the $\overline{\Sigma^{*0}}$ cross-section is equal to the averaged 
value of the \asigrp\ and \asigrm\ cross-sections.} and small 
fraction of the centrally produced \alam\ \cite{Malamcentr}. This 
gives 
\beq
[\axim/(\alam + \asigo)]_{prompt} = 0.14 \pm 0.04,~~~ [(\asigrm +
\asigrp)/2(\alam + \asigo)]_{prompt} = 0.14 \pm 0.03
\eeq
where we took into account that one can not separate the prompt 
\alam\ and $\overline{\Sigma^0}$. The JETSET and Pei models predict 
$[\xim/(\lam + \sigo)]_{prompt} = 0.11$ and 0.14, respectively, in 
good agreement with the Mirabelle result. 
For the $[\sigrp/(\lam + \sigo)]_{prompt}$, the same models 
predict 0.13 and 0.41, respectively. The  JETSET estimate is in good 
agreement with the Mirabelle result. The Pei model prediction is 
significantly higher, since the fractions of primary produced \lam\ 
and \sigo\ are smaller in this model
than in JETSET by the factors of 3.7 and 2.2, respectively.
Notice that good agreement of the Pei model with the Mirabelle 
result for the $[\xim/(\lam + \sigo)]_{prompt}$ and disagreement  
for the $[\sigrp/(\lam + \sigo)]_{prompt}$ is not contradictory, 
since the fraction of the promptly produced \xim\ in this model is 
also smaller by a factor of 1.8 than in JETSET.	One can also notice 
that the Pei model predicts essentially the same ratios \sigrp/\lam\ 
and \del/\prot\ for the promptly produced baryons. Therefore if the 
Mirabelle result for the $[\sigrp/(\lam + \sigo)]_{prompt}$ ratio  
is not biased and indeed represents a good estimate of the corresponding 
ratio at LEP, as indicated by all results of this paper, this implies 
that the fractions of promptly produced octet baryons in the Pei model 
might be underestimated. The quark-combinatorics \cite{as} predicts 
$[\xim/(\lam + \sigo)]_{prompt} = \lambda/2$ and 
$[\sigrp/(\lam + \sigo)]_{prompt} =1$. The former just tests the strangeness
suppression factor and agrees with the usually accepted $\lambda \approx 0.3$.
The latter is a factor of 7 higher than the Mirabelle value and deviates 
from it by 29 standard deviations.  

In conclusion, we have shown that a study of the 
vector-to-pseudoscalar ratios at LEP as a function of $x_p$ allows to 
obtain estimates of these ratios for promptly produced mesons at 
$x_p \rightarrow 1$. These estimates, as well as the 
\axim/(\alam + \asigo) and (\asigrm + \asigrp)/2(\alam + \asigo) ratios 
for the total rates, have been found in very good agreement with 
the results of the Mirabelle and BEBC \kpp\ experiments. This shows 
that the fragmentation properties of the $\bar s$ valence  quark of 
the incident kaon and strange quarks produced in \ee\ annihilation 
are very similar. In fact, this interesting experimental observation 
is not unexpected.
In the Lund String  model \cite{lund} (implemented in the JETSET 7.4 
generator), quark-antiquark pair in an $\ee \rightarrow q\bar q$ event 
is produced in a colour force field stretched between the $q$ and the 
$\bar q$. The hadron production is viewed as a breaking of the string 
which can be interpreted as virtual $q\bar q$ pair production in a 
colour flux-tube. A soft hadronic collision is also considered in 
this model as a colour separation mechanism whereby valence quarks 
of incident meson act as borders of colour strings, analogous to the 
$q - \bar q$ field in \ee\ annihilation. Therefore it is not 
surprising that many features in fragmentation of the $\bar s$ 
valence quark of the incident \kp\ and strange quarks in \ee\ 
annihilation must be similar. Other fragmentation models for soft, 
low-$p_T$ hadron-hadron interactions such as the Dual Sheet models 
based on the Dual Topological Unitarization (DTU) (see \cite{capella},
for example) differ from the Lund model in their prescriptions for the 
number and topology of colour strings and in the role attributed to 
valence quarks (as discussed, for example, in \cite{Mkvtokphi,Bkvtok}). 
Therefore, a comparison of meson induced reactions with \ee\ data 
might allow to get better understanding of the fate of meson valence 
quarks and to discriminate among different approaches. 
%It appears 
%that the results of this paper add evidence in favor of fragmentation 
%models of DTU-type, where one of the incident meson valence quarks 
%($u$-quark for the \kp) is assumed to be ``held back'' with low c.m. 
%momentum rather then allowed to participate in fragmentation on equal 
%footing with the $\bar s$ valence quark.
 
%This is consistent with results obtained from a study of
%two-particle inclusive reactions $\kpp \rightarrow \ph + \roz + X$ 
%and $\kpp \rightarrow \ph + \kl^{*+}(892) + X$ (with $x_F>0.2$) at 
%32 GeV/$c$ \cite{twores}. The cross-sections of these reactions 
%are comparable within errors whereas a much larger cross-section 
%is expected for the latter reaction if the kaon valence $u$-quark 
%is allowed to participate in fragmentation on equal footing with 
%the kaon valence $\bar s$-quark.

The predictions of the JETSET model have been found in good 
agreement with the results of the LEP and \kpp\ experiments 
on the vector-to-pseudoscalar ratios for the promptly produced 
mesons as well as with the \axim/(\alam + \asigo) and 
(\asigrm + \asigrp)/2(\alam + \asigo) ratios for the total 
and direct rates. The Pei model predictions are also in 
reasonable agreement with the data, apart from the \roz/\pis\ ratio
and significant disagreement with the Mirabelle result on the 
(\asigrm + \asigrp)/2(\alam + \asigo) ratio for the direct 
rates. This indicates that the Pei prediction about significantly 
smaller fraction of promptly produced octet baryons in comparison 
with JETSET might be too strong. The simplest spin models of
fragmentation based on spin counting and, in particular, the 
quark-combinatorics model of Anisovich et al. are incompatible 
with the results of the LEP and \kpp\ experiments.

\section*{Acknowledgements}
\vskip 3mm
It is a great pleasure to thank my colleagues in the Mirabelle, BEBC and 
DELPHI experiments with whom many results considered in this paper were
obtained and discussed.

\end{document}